\documentclass[runningheads]{llncs}
\usepackage{graphicx}
\usepackage{amsmath}
\usepackage{amssymb}
\usepackage{dsfont}
\usepackage{graphicx,subcaption}
\usepackage{booktabs}
\usepackage{algorithm}
\usepackage{algpseudocode}
\usepackage{algcompatible}
\usepackage{multirow}
\usepackage{booktabs}
\usepackage{makecell}
\usepackage{cite}
\usepackage{xcolor}

\usepackage[colorlinks=true,allcolors=blue]{hyperref}
\begin{document}
\title{Unsupervised Anomaly Detection in Medical Images with a Memory-augmented Multi-level Cross-attentional Masked Autoencoder
%\yu{Do we need multi-level in the title, given that in abstract/intro we named our framework as 'memory-augmented masked autoencoder (MemMC-MAE)'}
}
\author{
Yu Tian \inst{1},
Guansong Pang \inst{5},
Yuyuan Liu \inst{2},
Chong Wang \inst{2},
Yuanhong Chen \inst{2},
Fengbei Liu \inst{2},
Rajvinder Singh \inst{3},
Johan W Verjans \inst{2,3,4}, 
Mengyu Wang \inst{1},
and 
Gustavo Carneiro\inst{6}
}

\institute{%No Institute Given
Harvard Ophthalmology AI Lab, Harvard University. \and 
Australian Institute for Machine Learning, University of Adelaide
\and  Faculty of Health and Medical Sciences, University of Adelaide
\and
South Australian Health and Medical Research Institute
\and
 Singapore Management University
 \and
 Centre for Vision, Speech and Signal Processing, University of Surrey
    }
\maketitle              % typeset the header of the contribution
\vspace{-.5cm}
\begin{abstract}
Unsupervised anomaly detection (UAD) aims to find anomalous images by optimising a detector using a training set that contains only normal images.
UAD approaches can be based on reconstruction methods, self-supervised approaches, and Imagenet pre-trained models.
Reconstruction methods, which detect anomalies from image reconstruction errors, are advantageous because they do not rely on the design of problem-specific pretext tasks needed by self-supervised approaches, and on the unreliable translation of models pre-trained from non-medical datasets.  However, reconstruction methods may fail because they can have low reconstruction errors even for anomalous images.
In this paper, we introduce a new reconstruction-based UAD approach that addresses this low-reconstruction error issue for anomalous images. 
Our UAD approach, the memory-augmented multi-level cross-attentional masked autoencoder (MemMC-MAE), is a transformer-based approach, consisting of a novel memory-augmented self-attention operator for the encoder and a new multi-level cross-attention operator for the decoder. 
MemMC-MAE masks large parts of the input image during its reconstruction, reducing the risk that it will produce low reconstruction errors because anomalies are likely to be masked and cannot be reconstructed. However, when the anomaly is not masked, then the normal patterns stored in the encoder's memory combined with the decoder's multi-level cross-attention will constrain the accurate reconstruction of the anomaly.
We show that our method achieves SOTA anomaly detection and localisation on colonoscopy, pneumonia, and covid-19 chest x-ray datasets. 

\end{abstract}

\section{Introduction}

Detecting and localising anomalous findings in medical images (e.g., polyps, malignant tissues, etc.) are of vital importance~\cite{tian2019one,tian2020few,litjens2017survey,baur2020scale,fan2020pranet,lz2020computer,liu2021self,liu2021acpl,liu2021noisy,luo2023harvard,tian2021self,tian2022contrastive,liu2022translation,shi2023artifact,chen2022bomd}. 
Systems that can tackle these tasks are often formulated with a classifier trained with large-scale datasets annotated by experts. 
Obtaining such annotation is often challenging in real-world clinical datasets because the amount of normal images from healthy patients tend to overwhelm the amount of anomalous images.
Hence, to alleviate the challenges of collecting anomalous images and learning from class-imbalanced training sets, the field has developed unsupervised anomaly detection (UAD) models~\cite{tian2021constrained,chen2021deep} that are trained exclusively with normal images. 
Such UAD strategy benefits from the straightforward acquisition of training sets containing only normal images and the potential generalisability to unseen anomalies without collecting all possible anomalous sub-classes.

Current UAD methods learn a one-class classifier (OCC) using only normal/healthy training data, and detect anomalous/disease samples using the learned OCC~\cite{f-AnoGAN,seebock2019exploiting,gong2019memorizing,chen2021deep,liu2019photoshopping,venkataramanan2020attention,pang2019deep,li2021cutpaste,tian2021pixel}. 
UAD methods can be divided into: 1) reconstruction methods, 2) self-supervised approaches, and 3) Imagenet pre-trained models.
Reconstruction methods~\cite{f-AnoGAN,gong2019memorizing,chen2021deep,liu2019photoshopping,venkataramanan2020attention} are trained to accurately reconstruct normal images, exploring the assumption that the lack of anomalous images in the training set will prevent a low error reconstruction of an test image that contains an anomaly. 
However, this assumption is not met in general because reconstruction methods are indeed able to successfully reconstruct anomalous images, particularly when the anomaly is subtle.
Self-supervised approaches~\cite{tian2021constrained,tian2021self,sohn2020learning} train models using contrastive learning, where pretext tasks must be designed to emulate normal and anomalous image changes for each new anomaly detection problem.
Imagenet pre-trained models~\cite{reiss2021panda,defard2020padim} produce features to be used by OCC, but the translation of these models into medical image problems is not straightforward.
Reconstruction methods are able to circumvent the aforementioned challenges posed by self-supervised and Imagenet pre-trained UAD methods, and they can be trained with a relatively small amount of normal samples.  
However, their viability depends on an acceptable mitigation of the potentially low reconstruction error of anomalous test images.

In this paper, we introduce a new UAD reconstruction method, the Memory-augmented Multi-level Cross-attention Masked Autoencoder (MemMC-MAE), designed to address the low reconstruction error of anomalous test images.
MemMC-MAE is a transformer-based approach based on masked autoencoder (MAE)~\cite{he2021masked} with of a novel memory-augmented self-attention encoder and a new multi-level cross-attention  decoder. 
MemMC-MAE masks large parts of the input image during its reconstruction, and given that the likelihood of masking out an anomalous region is large, then it is unlikely that it will accurately reconstruct that anomalous region.
However, there is still the risk that the anomaly is not masked out, so in this case, the normal patterns stored in the encoder’s memory combined with the
correlation of multiple normal patterns in the image, utilised by the decoder’s multi-level cross-attention can explicitly constrain the accurate anomaly reconstruction to produce high reconstruction error (high anomaly score).
The encoder’s memory is also designed to address the MAE's long-range 'forgetting' issue~\cite{martins2021infty}, which can be harmful for UAD due to the poor reconstruction based on forgotten normality patterns and 'unwanted' generalisability to subtle anomalies during testing.
Our contributions are summarised as:
\begin{itemize}
\item To the best of our knowledge, this is the first memory-based UAD method that relies on MAE~\cite{he2021masked};
\item A new memory-augmented self-attention operator for our MAE transformer encoder to explicitly encode and memorise the normality patterns; and
\item A novel decoder architecture that uses the learned multi-level memory-augmented encoder information as prior features to a cross-attention operator. 
\end{itemize}
Our method achieves better anomaly detection and localisation accuracy than most competing approaches on the UAD benchmarks using the public Hyper-Kvasir colonoscopy dataset~\cite{borgli2020hyperkvasir}, pneumonia~\cite{kermany2018identifying} and Covid-X~\cite{wang2020covid} Chest X-ray (CXR) dataset.

% \indent 

\section{Method}

\begin{figure}[t!]
    \centering
    \vspace{-10pt}
    \includegraphics[width=0.7\textwidth]{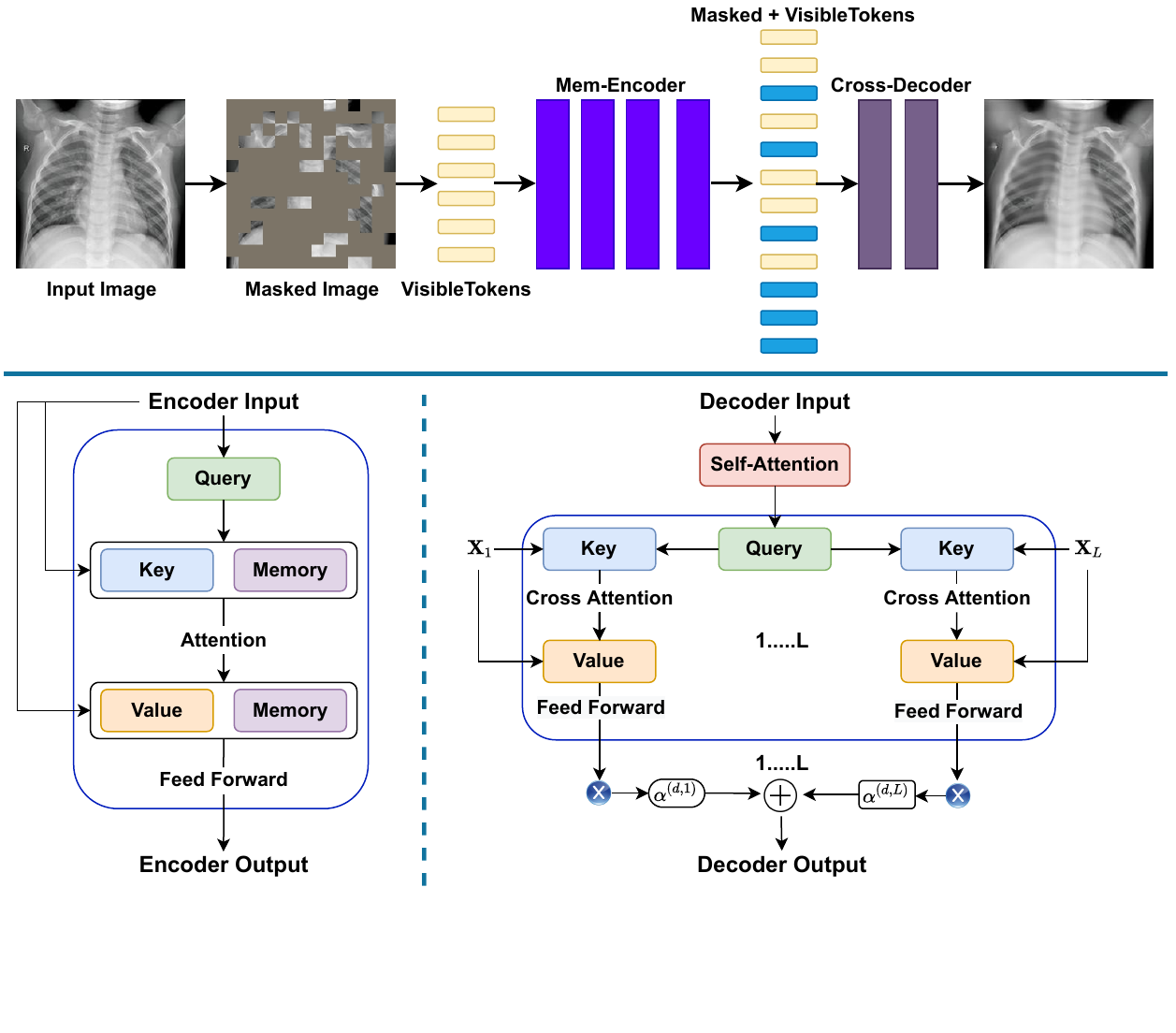}
    \vspace{-35pt}
    \caption{\textbf{Top:} overall MemMC-MAE framework. Yellow tokens indicate the unmasked visible patches, and blue tokens indicate the masked patches. Our memory-augmented transformer encoder only accepts the visible patches/tokens as input, and its output tokens are combined with dummy masked patches/tokens for the missing pixel reconstruction using our proposed multi-level cross-attentional transformer decoder.
    \textbf{Bottom-left:} proposed memory-augmented self-attention operator for the transformer encoder, and \textbf{bottom-right:} proposed multi-level cross-attention operator for the transformer decoder. 
    % \gustavo{Replace $\mathbb{X}$ by $\mathbf{X}$, and indices from 1 to L.  Also for $\alpha$, please write $\alpha^{(d,1)}$ and $\alpha^{(d,L)}$}.\yu{Sure. }
    }
    \label{fig:enc_dec_structure}
\end{figure}

%\subsection{Masked Autoencoders}

\subsection{Memory-augmented Multi-level Cross-attentional Masked Autoencoder (MemMC-MAE)}

%Masked Autoencoder~\cite{he2021masked} is a recently proposed self-supervised (SSL) pre-training approach for computer vision applications. 
%When training, MAE masks a high proportion of patches (75\%) from the input image and minimises the reconstruction error of the missing patches.
%Such training procedure is shown to be a meaningful SSL task, providing significant generalisation ability for multiple downstream tasks.
%Also, MAE has an asymmetric architecture, with a encoder that only takes the visible patches as input and a smaller/lighter decoder that reconstructs the original image based on the input tokens from visible and masked patches. 
%In this work, for the first time, 
% we argue that such task can be beneficial for \textit{training and testing} anomaly detectors because: 1) the high proportion of masked patches can eliminate anomalous patches from the input test image, forcing the network to reconstruct the masked patches from only the normal visible patches, and 2) learning such reconstruction task can allow the model to learn more effective  feature representations of normal samples than using Imagenet pre-trained models or contrastive learning with pretext tasks~\cite{defard2020padim,tian2021constrained,reiss2021panda}. 

Our MemMC-MAE, depicted in Fig.~\ref{fig:enc_dec_structure}, is based on the
masked autoencoder (MAE)~\cite{he2021masked} that was recently developed for the  pre-training of models to be used in downstream computer vision tasks. 
MAE has an asymmetric architecture, with a encoder that takes a small subset of the input image patches and a smaller/lighter decoder that reconstructs the original image based on the input tokens from visible patches and dummy tokens from masked patches. 
%We argue that MAE is beneficial for anomaly detection because: 1) the high proportion of masked patches can eliminate anomalous patches from the input test image, forcing the network to reconstruct the masked patches from only the normal visible patches, and 2) learning such reconstruction task can allow the model to learn more effective features of normal samples than using Imagenet pre-trained models or contrastive learning with pretext tasks~\cite{defard2020padim,tian2021constrained,reiss2021panda}. 

Our MemMC-MAE is trained with a normal image training set, denoted by $\mathcal{D} = \{ \mathbf{x}_i \}_{i=1}^{|\mathcal{D}|}$, where $\mathbf{x} \in \mathcal{X} \subset \mathbb{R}^{H \times W \times R}$ ($H$: height, $W$: width, $R$: number of colour channels). 
Our method first divides the input image $\mathbf{x}$ into non-overlapping patches $\mathcal{P} = \{ \mathbf{p}_i \}_{i=1}^{|\mathcal{P}|}$, where $\mathbf{p} \in  \mathbb{R}^{\hat{H} \times \hat{W} \times R}$, with $\hat{H}<<H$ and $\hat{W} << W$. We then randomly mask out 75\% of the $|\mathcal{P}|$ patches, and the remaining visible patches $\mathcal{P}^{(v)} = \{\mathbf{p}_{v}\}_{v=1}^{|\mathcal{P}^{(v)}|}$ (with $|\mathcal{P}^{(v)}| = 0.25\times |\mathcal{P}|$) are used by the MemMC-MAE to encode the normality patterns of those patches, and all $|\mathcal{P}^{(v)}|$ encoded visible patches and $|\mathcal{P}|-|\mathcal{P}^{(v)}|$ dummy masked patches are used as the input of a new multi-level cross-attention decoder to reconstruct the image.

The training of MemMC-MAE is based on the minimisation of the mean squared error (MSE) loss between the input and reconstructed images at the pixels of the masked patches of the training images.
The approach is evaluated on a testing set $\mathcal{T} = \{ (\mathbf{x},y,\mathbf{m})_i \}_{i=1}^{|\mathcal{T}|}$, where $y \in \mathcal{Y} = \{\text{normal}, \text{anomalous} \}$, and $\mathbf{m}\in \mathcal{M} \subset \{0,1\}^{H \times W \times 1}$ denotes the segmentation mask of the lesion in the image $\mathbf{x}$.
When testing, we also mask 75\% of the image and the patch-wise reconstruction error indicates anomaly localisation, and the mean reconstruction error of all patches is used to detect image-wise anomaly.
Below we provide details on the major contributions of MemMC-MAE, which are the memory-augmented transformer encoder that stores the long-term normality patterns of the training samples, and the new multi-level cross-attentional transformer decoder to leverage the correlation of features from the encoder to reconstruct the missing normal pixels. 

%Each encoder block contains a multi-head memory-augmented self-attention operator (see Fig.~\ref{fig:enc_dec_structure} - bottom left) and each decoder block contains a multi-level cross-attention operator (see Fig.~\ref{fig:enc_dec_structure} - bottom right). 

\subsubsection{Memory-augmented Transformer Encoder %\yu{Memory-augmented self-attention?}
(Fig.~\ref{fig:enc_dec_structure} - bottom left)}

%The main motivation of our memory-augmented transformer is the fact that reconstruction models can successfully reconstruct anomalous patches present in testing images, even when anomalous patches are not present in the training samples.Also, the training of transformer models can 'forget' the normal patches learned learned at early learning stages~\cite{martins2021infty}.

% Another motivation for the 
% memory-augmented transformer is the fact that AE models 
% can often successfully reconstruct abnormal patches present in testing images precisely because it did not learn such long-range relationship patterns between normal patches from the training images.

%often learn the patch relationships from abnormal testing patches (i.e., if those anomalous patches are not masked out) \gustavo{This is during testing, right? You don't access abnormal patches in training... needs clarification.}\yu{Yes during testing}, which often generalise too 'well' to even reconstruct the abnormal pixels \gustavo{during testing}. 

We modify the encoder from the transformer with our a novel memory-augmented self-attention, by extending the keys and values of the self-attention operation with learnable memory matrices that store normality patterns, which are updated via back-propagation. 
To this end, the proposed self-attention (SA) module for layer $l \in \{0,...,L-1\}$ is defined as: 
\begin{equation}
\begin{split}
\mathbf{X}^{(l+1)} &= 
f_{SA}\big(\mathbf{W}^{(l)}_{Q}\mathbf{X}^{(l)}, [\mathbf{W}^{(l)}_{K}\mathbf{X}^{(l)},\mathbf{M}^{(l)}_{K}], [\mathbf{W}^{(l)}_{V}\mathbf{X}^{(l)},\mathbf{M}^{(l)}_{V}] \big ), \\
%Mem(\mathcal{X}) &= Attention\big(\mathcal{W}_{Q}\mathcal{X}, Concat(\mathcal{W}_{K}\mathcal{X},M_{K}), Concat(\mathcal{W}_{V}\mathcal{X},M_{V}) \big )
\end{split}
    \label{eq:X}
\end{equation}
where $\mathbf{X}^{(0)}$ %\in \mathbb{R}^{D_{en} \times |\mathcal{P}^{(v)}|}$ 
is the encoder input matrix containing $|\mathcal{P}^{(v)}|$ patch tokens formed from the visible image patches transformed through the linear projection $\mathbf{W}^{(0)}$, %\in \mathbb{R}^{D_{en} \times \hat{H}\hat{W}R}$ 
with $|\mathcal{P}^{(v)}|$ being the number of visible tokens/patches, $\mathbf{X}^{(l)},\mathbf{X}^{(l+1)}$ %\in \mathbb{R}^{D_{en} \times |\mathcal{P}^{(v)}|}$ 
are the input and output of layer $l$, 
$\mathbf{W}^{(l)}_{Q},\mathbf{W}^{(l)}_{K},\mathbf{W}^{(l)}_{V}$  are the linear projections of the encoder's layer $l$ for query, key and value of the self-attention operator, respectively, and $\textbf{M}^{(l)}_{K},\textbf{M}^{(l)}_{V}$ are the layer $l$ learnable memory matrices that are concatenated with $\mathbf{W}_{K}\mathbf{X}^{(l)}$ and $\mathbf{W}_{V}\mathbf{X}^{(l)}$ using the operator $[.,.]$. 
%\yu{Do we really need the dimension for $\mathbf{W}^{(l)}_{Q},\mathbf{W}^{(l)}_{K},\mathbf{W}^{(l)}_{V}$, they are just MLP/linear layers. When I ready ViT and MAE original paper, they never mentioned the dimensions of those MLP weight matrix. Defining some new variable to represent the dimension of those weight matrix could looks too messy in the paper. } \gustavo{I added it there.  I think we need to add them for completeness.}
The self-attention operator $f_{SA}(.)$ follows the standard ViT~\cite{dosovitskiy2020image} and transformer~\cite{vaswani2017attention}, which computes a weighted sum of value vectors according to the cosine similarity distribution between query and key. 
Such memory-augmented self-attention aims to store normal patterns that are not encoded in the feature $\mathbf{X}^{(l)}$, %enforcing the encoder to store only normal features and 
forcing the decoder to reconstruct anomalous input patches into normal output patches during testing. 
%Our proposed memory-augmented self-attention are then adapted to the standard ViT transformer blocks to stack to the final memory-augmented transformer encoder. 

% \begin{figure}
%     \centering
%     \includegraphics[width=0.95\textwidth]{images/enc_dec_structure.pdf}
%     \caption{\textbf{Left} demonstrates the proposed memory-augmented self-attention for transformer encoder, and \textbf{right} shows the proposed cross-attention operator for transformer decoder.}
%     \label{fig:enc_dec_structure}
% \end{figure}

\subsubsection{Multi-level Cross-Attention Transformer Decoder (Fig.~\ref{fig:enc_dec_structure} - bottom right). 
%\yu{Prior driven multi-level cross-attention?}
}
Our transformer decoder computes the cross-attention operation using the outputs from all encoder layers and the decoder layer output from the self-attention operator (see Fig.~\ref{fig:enc_dec_structure} - Bottom right). More formally, the layer $d \in \{0,...,D-1\}$ of our decoder %cross-attention operator 
outputs
\begin{equation}
\begin{split}
\mathbf{Y}^{(d+1)} &= \sum_{l=1}^{L} \alpha^{(d,l)} \times f_{SA}\big(f_{SA}(\mathbf{Y}^{(d)},\mathbf{Y}^{(d)},\mathbf{Y}^{(d)}), \mathbf{W}^{(d)}_{K}\mathbf{X}^{(l)}, \mathbf{W}^{(d)}_{V}\mathbf{X}^{(l)}\big ),\\
%Dec(\mathbb{X}, \mathcal{Y}) &= \sum_{i=1}^{N} \alpha_{i} \times Attention\big(Attention(\mathcal{Y}), \mathcal{W}_{K}\mathbb{X}_{i}, \mathcal{W}_{V}\mathbb{X}_{i}\big )
\end{split}
    \label{eq:Y}
\end{equation}
where $\mathbf{Y}^{(d)}$ %\in \mathbb{R}^{N_{all} \times D_{d}}$ 
and $\mathbf{Y}^{(d+1)}$ %\in \mathbb{R}^{N_{all} \times D_{d}}$ 
represent the input and output of the decoder layer $d$ containing $|\mathcal{P}|$ tokens (i.e., $|\mathcal{P}^{(v)}|$ tokens from the visible patches of the encoder and $|\mathcal{P}| - |\mathcal{P}^{(v)}|$ dummy tokens from the masked patches), 
%, \gustavo{we need to say how $\mathbf{Y}^{(d)}$ is formed with the visible and masked tokens,} \yu{$\mathbf{Y}^{(d)}$ is formed with both visible and dummy masked tokens, and we add positional embedding to each of the token~\cite{he2021masked}. }
$\mathbf{X}^{(l)}$ denotes the output from encoder layer $l-1$, and $\mathbf{W}^{(d)}_{K},\mathbf{W}^{(d)}_{V}$ are the linear projections of the layer $d$ of the decoder for the key and value of the self-attention operator, respectively. Note that all $|\mathcal{P}|$ input tokens for the decoder are attached with positional embeddings.
%\gustavo{and the attention function $f_A(.)$ is defined as in~\cite{dosovitskiy2020image}, which does this and that}. 
The multi-level cross-attention results in~\eqref{eq:Y} are fused together with a weighted sum operation using the weight  $\alpha^{(l,d)}$, which is computed based on a linear projection layer and sigmoid function to control the weight of different layers' cross-attention results, as in 
\begin{equation}
\begin{split}
\alpha^{(d,l)} &= \sigma \left(\mathbf{W}_{\alpha}^{(d,l)} \left (\left [f_{SA}(\mathbf{Y}^{(d)},\mathbf{Y}^{(d)},\mathbf{Y}^{(d)}), \mathbf{Y}^{(d+1)}\right ] \right)\right), \\
%\alpha_{i} &= Sigmoid \big(FC \left (Concat (Attention(\mathcal{Y}), Dec(\mathbb{X}, \mathcal{Y})) \right)\big),
\end{split}
    \label{eq:alpha}
\end{equation}
where $\sigma(.)$ is the sigmoid function, and
$\mathbf{W}_{\alpha}^{(d,l)}$
%$FC$denotes the fully connected layer to produce
denotes a learnable weight matrix.
Such fusion mechanism enforces the correlation of multiple normal patterns in the image present at
different levels of encoding information to contribute at different decoding layers by adjusting their relative importance using the self-attention output
%$Attention(\mathcal{Y})$
from $f_{SA}(.)$ and cross-attention output 
$\mathbf{Y}^{(d+1)}$. %\gustavo{We'll need to discuss this in our meeting today.}
%$Dec(\mathbb{X}, \mathcal{Y})$.  

\subsection{Anomaly Detection and Segmentation}

We compute the anomaly score~\cite{chen2021deep} with multi-scale structural similarity (MS-SSIM)~\cite{wang2003multiscale}. 
The anomaly scores are pooled from 10 different random seeds for masking image patches with a fixed 75\% masking ratio, which enables a more robust anomaly detection and localisation. 
The anomaly localisation mask is obtained by computing the mean MS-SSIM scores for all patches, and the anomaly detection relies on the mean MS-SSIM scores from the patches~\cite{chen2021deep}.

\begin{table}[t]
\centering
\resizebox{0.66\textwidth}{!}{
\begin{tabular}{@{}@{\hskip .15in}c@{\hskip .15in}c@{\hskip .15in}c@{\hskip .15in}c@{\hskip .15in}c@{\hskip .15in}@{}}
\toprule \hline
Methods         & Publication  & Pneumonia & Covid-X   & Hyper-Kvasir \\ \hline\hline
DAE~\cite{masci2011stacked}     &  ICANN'11    & 0.599   & 0.557   & 0.705  \\
OCGAN~\cite{perera2019ocgan}   &  CVPR'18  &    0.703       & 0.612& 0.813  \\
F-anoGAN~\cite{f-AnoGAN}    &  IPMI'17     & 0.755 & 0.669  & 0.907    \\
ADGAN~\cite{liu2019photoshopping}    &  ISBI'19  &   0.627      & 0.659  & 0.913 \\
MS-SSIM~\cite{chen2021deep}   & AAAI'22     &  0.695  & 0.634   & 0.917      \\
PANDA~\cite{reiss2021panda}  & CVPR'21    & 0.657 & 0.629 & 0.937  \\  
PaDiM~\cite{defard2020padim}   & ICPR'21  & 0.663 & 0.614  & 0.923 \\ 
IGD~\cite{chen2021deep}     & AAAI'22     &  0.734  & 0.699   & 0.939   \\
CCD+IGD*~\cite{tian2021constrained}    & MICCAI'21  &   0.775     & 0.746   & \textbf{0.972}    \\ \hline
Ours       &   & \textbf{0.879}    & \textbf{0.917} &\textbf{0.972}    \\ 
 \hline \bottomrule
\end{tabular}
}
\caption{\textbf{Anomaly detection AUC} test results on Pneumonia and Covid-X Chest X-ray datasets and Hyper-Kvasir colonoscopy dataset. CCD+IGD*~\cite{tian2021constrained} requires at least $2\times$longer training time than other approaches in the table because of a two-stage self-supervised pre-training and fine-tuning.}
\vspace{-20pt}
\label{tab:detection_auc}
\end{table}

\section{Experiments and Results}

\subsubsection{Datasets and Evaluation Measures:}

Three disease screening datasets 
are used in our experiments. 
We test anomaly detection on the CXR images of the pneumonia chest X-ray dataset~\cite{kermany2018identifying} and Covid-X dataset~\cite{wang2020covid}, and both anomaly detection and localisation on the colonoscopy images of the Hyper-Kvasir dataset~\cite{borgli2020hyperkvasir}.
The publicly available \textbf{pneumonia chest X-ray dataset}~\cite{kermany2018identifying}, consisting of normal and pneumonia-affected images, was obtained from a total of 6,480 patients. In accordance with~\cite{zhao2021anomaly}, we structured the anomaly detection dataset such that the training set encompasses 1,349 normal images, and the testing set comprises 234 normal and 390 pneumonia images. Each chest X-ray image has been resized to the standardized dimensions of 256 $\times$ 256 pixels.
\textbf{Covid-X}~\cite{wang2020covid} has a training set with 1,670 Covid-19 positive and 13,794 Covid-19 negative CXR images, but we only use the 13,794 Covid-19 negative CXR images for training. The test set contains 400 CXR images, consisting of 200 positive and 200 negative images, each image with size 299 $\times$ 299 pixels. 
\textbf{Hyper-Kvasir} is a large-scale public gastrointestinal dataset. The images were collected from the gastroscopy and colonoscopy procedures from Baerum Hospital in Norway, and were annotated by experienced medical practitioners. The dataset contains 110,079 images from unhealthy and healthy patients, out of which, 10,662 are labelled. 
%We use part of the labelled images from the dataset to train our MemMC-MAE.
Following~\cite{tian2021constrained}, 2,100 normal images are selected, from which we use 1,600 for training and 500 for testing. The testing set also contains 1,000 anomalous images with their segmentation masks. 
Detection is assessed with area under the ROC curve (AUC), and localisation is evaluated with intersection over union (IoU).

\subsubsection{Implementation Details}

For the transformer, 
we follow ViT-B~\cite{dosovitskiy2020image,he2021masked} for designing the encoder and decoder, consisting of stacks of transformer blocks. 
Inspired by U-Net~\cite{zhou2018unet++} for medical segmentation, we add  residual connections to transfer information from earlier to later blocks for both the encoder and decoder. 
Each encoder block contains a memory-augmented self-attention block and an MLP block with LayerNorm (LN). Each decoder block contains a multi-level cross-attention block and an MLP block with LayerNorm (LN).
We also adopt a linear projection layer after the
encoder to match the different width between encoder and decoder~\cite{he2021masked}. 
We add positional embeddings (with the sine-cosine version) to both the encoder and decoder input tokens. 
RandomResizedCrop is used for data augmentation during training. 
Our method is trained for 2000 epochs in an end-to-end manner using the Adam optimiser with a weight decay of 0.05 and a batch size of 256. 
The learning rate is set to 1.5e-3. In the beginning, we warm up the training process for 5 epochs. The method is implemented in PyTorch and runs on an NVIDIA 3090 GPU. The overall training time is around 22 hours, and the mean inference time takes 0.21s per image. 

\subsubsection{Evaluation on Anomaly Detection on Chest X-ray and Colonoscopy} 
We compare our method with nine competing UAD approaches:
DAE~\cite{masci2011stacked}, OCGAN~\cite{perera2019ocgan}, f-anogan~\cite{f-AnoGAN}, ADGAN~\cite{liu2019photoshopping}, MS-SSIM autoencoder~\cite{chen2021deep}, PANDA~\cite{reiss2021panda}, PaDiM~\cite{defard2020padim}, CCD~\cite{tian2021constrained} and IGD~\cite{chen2021deep}. 
We apply the
same experimental setup (i.e., image pre-processing, training strategy, evaluation methods) to these methods above as the one for our approach for fair comparison.
% \noindent\textbf{Evaluation Measures.}
% Similarly to previous papers~\cite{masci2011stacked,gong2019memorizing,perera2019ocgan,chen2021deep}, we use the area under the ROC curve (AUC) as the evaluation measure for anomaly detection, where
% larger AUC values suggest better performance.
%\noindent\textbf{Detection results on Covid-X and Hyper-Kvasir.} 
The quantitative comparison results for anomaly detection are shown in Table~\ref{tab:detection_auc} for Pneumonia, Covid-X, and Hyper-Kvasir benchmarks. Our MemMC-MAE achieves the best AUC results on three datasets with 87.9\%, 91.7\% and 97.2\%, respectively.
On pneumonia chest x-ray dataset, our model surpasses the previous SOTA approaches by a minimum 10.4\% AUC and a maximum 28\% AUC. 
On Covid-X, our result outperforms all competing methods by a large margin with an improvement of 17.1\% over the second best approach. For Hyper-Kvasir, our result is on par with the best result in the field produced by CCD+IGD~\cite{tian2021constrained}, which has a training time $2\times$ longer than our approach.

%. We obtain AUC improvements of 17.1\%  and 3.3\% over the second best approach, on Covid-X and Hyper-Kvasir, respectively. 

\begin{figure}[!t]
    \centering
    \includegraphics[width=0.8\textwidth]{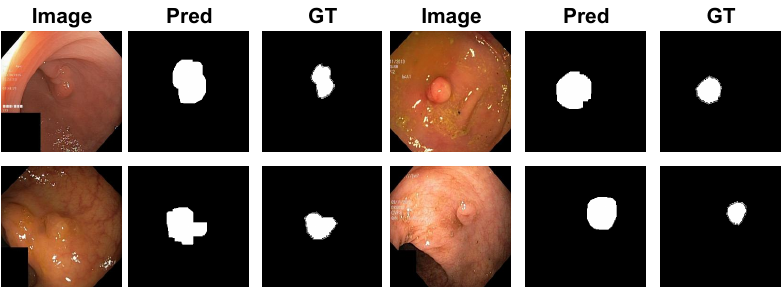}
    \vspace{-10pt}
    \caption{Segmentation results of our proposed method on Hyper-Kvasir~\cite{borgli2020hyperkvasir}, with our predictions (Pred) and ground truth annotations (GT).  
    }
    \label{fig:qualitative_segmentation}
    \vspace{-10pt}
\end{figure}

\begin{figure}[t]
    \centering
    \includegraphics[width=1\textwidth]{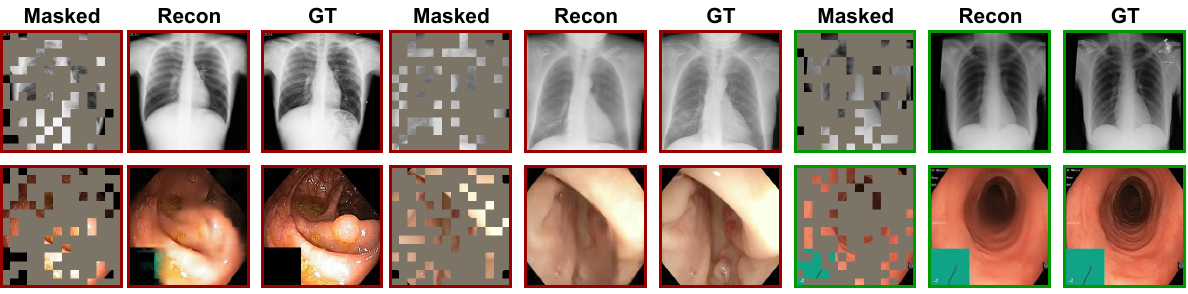}
    \vspace{-15pt}
    \caption{Reconstruction of testing images from Covid-X (Top) and Hyper-Kvasir (Bottom). For each triplet, we show the masked image (left), our MemMC-MAE reconstruction (middle), and the ground-truth (right). Normal testing images are marked with green boxes, and anomalous ones are marked with red boxes.
    }
    \label{fig:qualitative}
    
\end{figure}

\begin{table}[t]
\scalebox{.9}{
\parbox{.55\linewidth}{
\centering
% \resizebox{\linewidth}{!}{%
\begin{tabular}{ccc|cc}
\toprule\hline
MAE & Mem-Enc & MC-Dec   &  AUC - Covid &  AUC - Hyper \\ \hline \hline
\checkmark       &           &                      & 0.799  & 0.915          \\
  \checkmark      & \checkmark         &               &     0.862            & 0.956     \\\hline
   \checkmark     & \checkmark   & \checkmark          &      \textbf{0.917}  & \textbf{0.972}  \\        \hline\bottomrule
\end{tabular}%
\caption{\textbf{Ablation study} on Covid-X of the encoder's memory-augmented operator (Mem-Enc) and the decoder's multi-level cross-attention (MC-Dec).}
\label{tab:ablation}
}
\hfill
% \vspace{.2in}
\parbox{.55\linewidth}{
\centering
\begin{tabular}{@{}cc@{}}
\toprule \hline
Methods & Localisation - IoU   \\ \hline \hline
IGD~\cite{chen2021deep}     & 0.276  \\
PaDiM~\cite{defard2020padim}   & 0.341 \\
CAVGA-$R_{u}$~\cite{venkataramanan2020attention}     & 0.349  \\
CCD + IGD~\cite{tian2021constrained}        &  0.372 \\
% CCD + PaDiM~\cite{tian2021constrained}        &  0.341 \\
\hline
Ours        &  \textbf{0.419} \\\bottomrule \hline
\end{tabular}%
\caption{\textbf{Anomaly localisation:} Mean IoU test results on Hyper-Kvasir on 5 groups of 100 images.}
% \vspace{-.25in}
\label{tab:localisation_auc_HK}
}
}
\vspace{-10pt}
\end{table}

\subsubsection{Evaluation on Anomaly Localisation on Colonoscopy} 
%\noindent\textbf{Baselines.}  
We compare our anomaly localisation results on Table~\ref{tab:localisation_auc_HK} with four recently proposed UAD baselines: IGD~\cite{chen2021deep}, PaDiM~\cite{defard2020padim}, CCD~\cite{tian2021constrained} and CAVGA-$R_{u}$~\cite{venkataramanan2020attention}. 
The results of these methods on Table~\ref{tab:localisation_auc_HK} are from~\cite{tian2021constrained}. 
% \gustavo{Yu, is this correct? Were they taken from~\cite{tian2021constrained}?}\yu{Yes.}
% \noindent\textbf{Evaluation Measures.} 
% Follow to previous papers~\cite{tian2021constrained}, for anomaly localisation, the performance is measured by Intersection over Union (IoU). Larger IoU values indicate better performance. 
%\noindent\textbf{Localisation results on Hyper-Kvasir.}
%Table shows the results on Hyper-Kvasir~\ref{tab:localisation_auc_HK}. 
Following~\cite{tian2021constrained}, we randomly sample five groups of 100 anomalous images from the test set and compute the mean segmentation IoU.
The proposed MemMC-MAE surpasses IGD, PaDiM, CAVGA-$R_{u}$ and CCD by a minimum of 4.7\% and a maximum of 14.3\% IoU, illustrating the effectiveness of our model in localising anomalous tissues. 

%\subsection{Qualitative Results}
\subsubsection{Visualisation of predicted segmentation.}
The visualisation of polyp segmentation results of MemMC-MAE on Hyper-Kvasir~\cite{borgli2020hyperkvasir} is shown in Fig.~\ref{fig:qualitative_segmentation}. Notice that our model can accurately segment colon polyps of various sizes and shapes.

\subsubsection{Visualisation of Reconstructed Images}
Figure~\ref{fig:qualitative} shows the reconstructions produced by MemMC-MAE on Covid-X (Top) and Hyper-Kvasir (Bottom) testing images. 
Notice that our method can effectively reconstruct the anomalous images with
polyps/covid as normal images by automatically removing the polyps or blurring the anomalous regions, leading to larger reconstruction errors for those anomalies. 
The normal images are accurately reconstructed with smaller reconstruction errors than the anomalous images. 
%Such reconstruction error deviations indicate the effectiveness of our approach on detecting malignant lesions. 

\subsubsection{Ablation Study}
Tab.~\ref{tab:ablation} shows the contribution of each component of our proposed method on Covid-X and Hyper-Kvasir testing set. The baseline MAE~\cite{he2021masked} achieves 79.9\% and 91.5\% AUC on the two datasets, respectively.
Our method obtains a significant performance gain by adding the memory-augmented self-attention operator to the transformer encoder (Mem-Enc). 
Adding the proposed multi-level cross-attention operator into the decoder (MC-Dec) further boosts the performance on both datasets.

\section{Conclusion}
% In general, our work aims to brought broader impacts to the community that a simple bert-like training/pre-training can provide significant improvements to UAD problems. 
We proposed a new UAD reconstruction method, called MemMC-MAE, for anomaly detection and localisation in medical images, which to the best of our knowledge, is the first memory-based UAD method using MAE. 
MemMC-MAE introduced a novel memory-augmented self-attention operator for the MAE encoder and a new multi-level cross-attention for the MAE decoder to address the large reconstruction error of anomalous images that plague UAD reconstruction methods. 
The resulting anomaly detector showed SOTA anomaly detection and localisation accuracy on three public medical datasets.
Despite the remarkable performance, the results can potentially improve if we use MemMC-MAE as a pre-training approach for other UAD methods, which we plan to explore in the future.
\bibliographystyle{splncs04}
\bibliography{mybibliography}

\begin{thebibliography}{10}
\providecommand{\url}[1]{\texttt{#1}}
\providecommand{\urlprefix}{URL }
\providecommand{\doi}[1]{https://doi.org/#1}

\bibitem{baur2020scale}
Baur, C., et~al.: Scale-space autoencoders for unsupervised anomaly
  segmentation in brain mri. In: MICCAI. pp. 552--561. Springer (2020)

\bibitem{borgli2020hyperkvasir}
Borgli, H., et~al.: Hyperkvasir, a comprehensive multi-class image and video
  dataset for gastrointestinal endoscopy. Scientific Data  \textbf{7}(1),
  1--14 (2020)

\bibitem{chen2021deep}
Chen, Y., Tian, Y., Pang, G., Carneiro, G.: Deep one-class classification via
  interpolated gaussian descriptor. arXiv preprint arXiv:2101.10043  (2021)

\bibitem{chen2022bomd}
Chen, Y., et~al.: Bomd: Bag of multi-label descriptors for noisy chest x-ray
  classification. arXiv preprint arXiv:2203.01937  (2022)

\bibitem{defard2020padim}
Defard, T., Setkov, A., Loesch, A., Audigier, R.: Padim: a patch distribution
  modeling framework for anomaly detection and localization. arXiv preprint
  arXiv:2011.08785  (2020)

\bibitem{dosovitskiy2020image}
Dosovitskiy, A., et~al.: An image is worth 16x16 words: Transformers for image
  recognition at scale. arXiv preprint arXiv:2010.11929  (2020)

\bibitem{fan2020pranet}
Fan, D.P., Ji, G.P., Zhou, T., Chen, G., Fu, H., Shen, J., Shao, L.: Pranet:
  Parallel reverse attention network for polyp segmentation. In: MICCAI. pp.
  263--273. Springer (2020)

\bibitem{gong2019memorizing}
Gong, D., et~al.: Memorizing normality to detect anomaly: Memory-augmented deep
  autoencoder for unsupervised anomaly detection. In: ICCV. pp. 1705--1714
  (2019)

\bibitem{he2021masked}
He, K., Chen, X., Xie, S., Li, Y., Doll{\'a}r, P., Girshick, R.: Masked
  autoencoders are scalable vision learners. arXiv preprint arXiv:2111.06377
  (2021)

\bibitem{kermany2018identifying}
Kermany, D.S., Goldbaum, M., Cai, W., Valentim, C.C., Liang, H., Baxter, S.L.,
  McKeown, A., Yang, G., Wu, X., Yan, F., et~al.: Identifying medical diagnoses
  and treatable diseases by image-based deep learning. cell  \textbf{172}(5),
  1122--1131 (2018)

\bibitem{li2021cutpaste}
Li, C.L., et~al.: Cutpaste: Self-supervised learning for anomaly detection and
  localization. In: CVPR. pp. 9664--9674 (2021)

\bibitem{litjens2017survey}
Litjens, G., et~al.: A survey on deep learning in medical image analysis.
  Medical image analysis  \textbf{42},  60--88 (2017)

\bibitem{liu2021noisy}
Liu, F., Tian, Y., Cordeiro, F.R., Belagiannis, V., Reid, I., Carneiro, G.:
  Noisy label learning for large-scale medical image classification. arXiv
  preprint arXiv:2103.04053  (2021)

\bibitem{liu2021self}
Liu, F., Tian, Y., et~al.: Self-supervised mean teacher for semi-supervised
  chest x-ray classification. arXiv preprint arXiv:2103.03629  (2021)

\bibitem{liu2021acpl}
Liu, F., et~al.: Acpl: Anti-curriculum pseudo-labelling for semi-supervised
  medical image classification. CVPR  (2022)

\bibitem{liu2019photoshopping}
{Liu}, Y., et~al.: Photoshopping colonoscopy video frames. In: ISBI. pp.~1--5
  (2020)

\bibitem{liu2022translation}
Liu, Y., Tian, Y., Wang, C., Chen, Y., Liu, F., Belagiannis, V., Carneiro, G.:
  Translation consistent semi-supervised segmentation for 3d medical images.
  arXiv preprint arXiv:2203.14523  (2022)

\bibitem{luo2023harvard}
Luo, Y., et~al.: Harvard glaucoma fairness: A retinal nerve disease dataset for
  fairness learning and fair identity normalization. arXiv preprint
  arXiv:2306.09264  (2023)

\bibitem{lz2020computer}
LZ, C.T.P., et~al.: Computer-aided diagnosis for characterisation of colorectal
  lesions: a comprehensive software including serrated lesions.
  Gastrointestinal Endoscopy  (2020)

\bibitem{martins2021infty}
Martins, P.H., Marinho, Z., Martins, A.F.: infinity-former: Infinite memory
  transformer. arXiv preprint arXiv:2109.00301  (2021)

\bibitem{masci2011stacked}
Masci, J., et~al.: Stacked convolutional auto-encoders for hierarchical feature
  extraction. In: ICANN. pp. 52--59. Springer (2011)

\bibitem{pang2019deep}
Pang, G., Shen, C., van~den Hengel, A.: Deep anomaly detection with deviation
  networks. In: Proceedings of the 25th ACM SIGKDD International Conference on
  Knowledge Discovery \& Data Mining. pp. 353--362 (2019)

\bibitem{perera2019ocgan}
Perera, P., Nallapati, R., Xiang, B.: Ocgan: One-class novelty detection using
  gans with constrained latent representations. In: CVPR. pp. 2898--2906 (2019)

\bibitem{reiss2021panda}
Reiss, T., Cohen, N., Bergman, L., Hoshen, Y.: Panda: Adapting pretrained
  features for anomaly detection and segmentation. In: Proceedings of the
  IEEE/CVF Conference on Computer Vision and Pattern Recognition. pp.
  2806--2814 (2021)

\bibitem{f-AnoGAN}
Schlegl, T., et~al.: f-anogan: Fast unsupervised anomaly detection with
  generative adversarial networks. Medical image analysis  \textbf{54},  30--44
  (2019)

\bibitem{seebock2019exploiting}
Seeb{\"o}ck, P., et~al.: Exploiting epistemic uncertainty of anatomy
  segmentation for anomaly detection in retinal oct. IEEE transactions on
  medical imaging  \textbf{39}(1),  87--98 (2019)

\bibitem{shi2023artifact}
Shi, M., et~al.: Artifact-tolerant clustering-guided contrastive embedding
  learning for ophthalmic images in glaucoma. IEEE Journal of Biomedical and
  Health Informatics  (2023)

\bibitem{sohn2020learning}
Sohn, K., Li, C.L., Yoon, J., Jin, M., Pfister, T.: Learning and evaluating
  representations for deep one-class classification. arXiv preprint
  arXiv:2011.02578  (2020)

\bibitem{tian2020few}
Tian, Y., Maicas, G., Pu, L.Z.C.T., Singh, R., Verjans, J.W., Carneiro, G.:
  Few-shot anomaly detection for polyp frames from colonoscopy. In: MICCAI. pp.
  274--284. Springer (2020)

\bibitem{tian2022contrastive}
Tian, Y., Pang, G., Liu, F., Liu, Y., Wang, C., Chen, Y., Verjans, J.,
  Carneiro, G.: Contrastive transformer-based multiple instance learning for
  weakly supervised polyp frame detection. In: MICCAI. pp. 88--98. Springer
  (2022)

\bibitem{tian2021constrained}
Tian, Y., Pang, G., Liu, F., Shin, S.H., Verjans, J.W., Singh, R., Carneiro,
  G., et~al.: Constrained contrastive distribution learning for unsupervised
  anomaly detection and localisation in medical images. MICCAI 2021  (2021)

\bibitem{tian2019one}
Tian, Y., et~al.: One-stage five-class polyp detection and classification. In:
  2019 IEEE 16th International Symposium on Biomedical Imaging (ISBI 2019). pp.
  70--73. IEEE (2019)

\bibitem{tian2021pixel}
Tian, Y., et~al.: Pixel-wise energy-biased abstention learning for anomaly
  segmentation on complex urban driving scenes. arXiv preprint arXiv:2111.12264
   (2021)

\bibitem{tian2021self}
Tian, Y., et~al.: Self-supervised multi-class pre-training for unsupervised
  anomaly detection and segmentation in medical images. arXiv preprint
  arXiv:2109.01303  (2021)

\bibitem{vaswani2017attention}
Vaswani, A., Shazeer, N., Parmar, N., Uszkoreit, J., Jones, L., Gomez, A.N.,
  Kaiser, {\L}., Polosukhin, I.: Attention is all you need. Advances in neural
  information processing systems  \textbf{30} (2017)

\bibitem{venkataramanan2020attention}
Venkataramanan, S., Peng, K.C., Singh, R.V., Mahalanobis, A.: Attention guided
  anomaly localization in images. In: ECCV. pp. 485--503. Springer (2020)

\bibitem{wang2020covid}
Wang, L., Lin, Z.Q., Wong, A.: Covid-net: A tailored deep convolutional neural
  network design for detection of covid-19 cases from chest x-ray images.
  Scientific Reports  \textbf{10}(1),  1--12 (2020)

\bibitem{wang2003multiscale}
Wang, Z., et~al.: Multiscale structural similarity for image quality
  assessment. In: The Thrity-Seventh Asilomar Conference on Signals, Systems \&
  Computers, 2003. vol.~2, pp. 1398--1402. Ieee (2003)

\bibitem{zhao2021anomaly}
Zhao, H., Li, Y., He, N., Ma, K., Fang, L., Li, H., Zheng, Y.: Anomaly
  detection for medical images using self-supervised and translation-consistent
  features. IEEE Transactions on Medical Imaging  \textbf{40}(12),  3641--3651
  (2021)

\bibitem{zhou2018unet++}
Zhou, Z., et~al.: Unet++: A nested u-net architecture for medical image
  segmentation. In: Deep learning in medical image analysis and multimodal
  learning for clinical decision support, pp. 3--11. Springer (2018)

\end{thebibliography}




% LNCS Homepage, \url{http://www.springer.com/lncs}. Last accessed 4
% \end{thebibliography}
\end{document}